\documentclass[twocolumn,aps,amsmath,amssymb,showpacs,superscriptaddress]{revtex4}

\usepackage{graphicx}
\usepackage{dcolumn}
\usepackage{bm}

\begin{document}


\title{Effective interaction and condensation of dipolaritons in coupled quantum wells}
\author{Tim Byrnes}
\affiliation{National Institute of Informatics, 2-1-2
Hitotsubashi, Chiyoda-ku, Tokyo 101-8430, Japan}

\author{German V. Kolmakov}
\affiliation{Physics Department, New York City College of Technology, The City University of New York, Brooklyn, New York 11201, USA}

\author{Roman Ya. Kezerashvili}
\affiliation{Physics Department, New York City College of Technology, The City University of New York, Brooklyn, New York 11201, USA}
\affiliation{The Graduate School and University Center, The City University of New York, New York, New York 10016, USA}

\author{Yoshihisa Yamamoto}
\affiliation{National Institute of Informatics, 2-1-2
Hitotsubashi, Chiyoda-ku, Tokyo 101-8430, Japan}
 \affiliation{E.
L. Ginzton Laboratory, Stanford University, Stanford, CA 94305}

\date{\today}

\date{\today}

\begin{abstract}
Dipolaritons are a three-way superposition of photon, a direct exciton, and an indirect exciton that are formed in coupled quantum well microcavities.  
As is the case with exciton-polaritons, dipolaritons have a self-interaction due to direct and exchange effects of the underlying electrons and holes.  
Here we present a theoretical description of dipolaritons and derive simple formulas for their basic parameters. In particular, we derive the effective dipolariton-dipolariton interaction taking into account of exchange effects between the excitons.  We obtain a simple relation to describe the effective interaction at low densities.  We find that dipolaritons should condense under suitable conditions, described by a dissipative Gross-Pitaevskii equation. While the parameters for condensation are promising, we find that the level of tunability of the interactions is limited.  
\end{abstract}

\pacs{71.36.+c, 71.35.-y, 03.67.Ac}
\maketitle

\section{Introduction}

The observation of the condensation of exciton-polaritons \cite{deng02,kasprzak06,balili07} has created a large amount of interest in the last decade \cite{deng10,keeling11,carusotto13}. 
Exciton-polariton condensates display fascinating properties such as superfluidity \cite{amo09,amo09b}, vortex formation \cite{lagoudakis08,lagoudakis09,roumpos11}, 
and has been suggested for use in future technologies such as polaritronics, the polariton analogue of atomtronics \cite{deveaud08,amo10,ballarini13}, quantum simulators \cite{kim11,byrnes10,tanese13}, and novel light sources \cite{imamoglu96,byrnes13}. One recent development is the observation of dipolaritons -- bosonic quasiparticles formed in coupled double quantum wells embedded into a  microcavity formed by two distributed Bragg reflectors (Fig. 1) \cite{cristofolini12}. Compared to the exciton-polariton, which is a quasiparticle consisting of a superposition of a photon and an exciton, a dipolariton is a three-way superposition of a microcavity photon, direct exciton (DX), and an indirect exciton (IX).  The DX is a bound electron-hole pair in the same quantum well, and an IX is a bound electron-hole pair between the quantum wells.  The coupling to photons means that 
it is a new type of polariton, which shares similar properties to the exciton-polariton such as light effective mass, but in addition has a dipole moment \cite{kyriienko13}.  The dipole moment is expected to enhance the dipolariton-dipolariton interactions, but in a tunable way by varying the relative proportions of the photon, DX, and IX. This is interesting from a quantum optoelectronic
standpoint with potential applications to coherent transfer between photons to electrons \cite{cristofolini12}.  Although condensation of dipolaritons has not been observed to date, the light effective mass of the dipolaritons suggest that the prospect of this is rather promising.  Despite the experimental interest, currently there is no rigorous theoretical treatment of the properties of the dipolariton, in particular the effective dipolariton-dipolariton interaction.

\begin{figure}
\scalebox{0.65}{\includegraphics{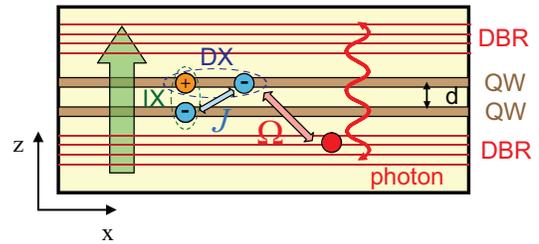}}
\caption{\label{fig:device} (Color online) 
The schematic device configuration considered in this paper. A coupled double quantum well (QW) semiconductor microstructure is sandwiched by two distributed Bragg reflectors (DBRs). On the application of an electric field (large arrow) this structure allows for 
three types of bare quasiparticles: the indirect exciton (IX) consisting of an electron and hole in each QW, a direct exciton (DX) consisting of the electron and hole being in the same quantum well, and a microcavity photon. A dipolariton consists of a superposition of the three particle species due to the coherent couplings as marked. }
\end{figure}

Several works have in the past have calculated
the effective exciton-exciton interaction \cite{bobrysheva72,hanamura79,stolz81,schmittrink85,cuiti98,rochat00,deleon01}, from which the 
polariton-polariton interaction can be obtained simply by multiplying by the exciton fraction.  The exciton-exciton interaction originate from the Coulomb interaction of the underlying electron and holes making up the excitons. The interaction is typically described as the sum of two 
contributions -- the ``direct'' and ``exchange'' contributions \cite{cuiti98}.  The direct contribution corresponds to exciton-exciton scattering process 
$ (e,h) + (e',h') \rightarrow (e,h) + (e',h') $, where $ (e,h) $ denotes an exciton containing an electron $ e $ and a hole $ h $.  The dashed and undashed labels refer to wavefunction coordinate labels in first quantized formalism, where the total wavefunction is antisymmetrized with respect to the electrons and holes exchange. The exchange contribution corresponds to the exchange exciton-exciton scattering $ (e,h) + (e',h') \rightarrow (e,h') + (e',h) $, where one of the underlying fermions is exchanged.  It is well-known that under typical 
densities the dominant process for DXs is the exchange contribution, giving the standard interaction $ \hbar g \approx   \frac{6e^2}{4 \pi \epsilon a_B}  \frac{a_B^2}{A} $, where $ a_B $ is the Bohr radius of the exciton, $ A $ is the sample area, $ e $ is the charge of the electron, and $ \epsilon $ is the permittivity of the semiconductor.  In Ref. \cite{byrnes10}, this procedure was generalized to 
IXs.  Here it was found that for a non-zero quantum well separation (see Fig. \ref{fig:device}), both the direct and exchange contributions need to be taken into account to obtain the effective IX-IX interaction.  

In this paper we present a theoretical description of dipolaritons within coupled double quantum wells in semiconductor microcavity structures.  In particular we give a detailed derivation of the effective dipolariton-dipolariton interaction. Due to the three-way superposition, the interaction
will result from the total of DX-DX interaction, IX-IX interaction, and the IX-DX interaction. While the DX-DX and IX-IX interaction were analyzed before, here we give a detailed calculation of the 
IX-DX interaction, which has not appeared in the literature before.  Each of the three contributions to the interactions have a direct and exchange contribution, and we will show that it is important to include exchange effects for the IX-DX interaction. After performing the full calculation, we are able to obtain a simple expression for the dipolariton-dipolariton interaction valid for low dipolariton densities.  For readers that are uninterested in the details, the primary result of this paper is Eq. (\ref{mainresult}), which gives the effective dipolariton-dipolariton interaction. As mentioned above, due to the light dipolariton mass, they are promising for realization of condensation.  We give simple formulas for parameters that would describe the condensate, modeled in terms of a dissipative Gross-Pitaevskii equation \cite{wouters07}, which is the standard way to describe polariton condensates.  We give estimates of parameters and show the suitability of dipolaritons for condensation. 

This paper is organized as follows.  In Sec. \ref{sec:dipolham} we  derive the dipolariton Hamiltonian starting from the photon, direct exciton, and indirect exciton constituents.  This will serve to identify what quantities are necessary to calculate regarding the various interaction contributions between the constituent species.  
In Sec. \ref{sec:eff} we discuss the DX-DX, IX-IX, and IX-DX interactions.  The IX-DX interaction is calculated in detail while the known results for the DX-DX and IX-IX interaction are quoted for convenience.   In Sec. \ref{sec:gpequation} we present the dissipative Gross-Pitaevskii equation for dipolaritons, along with simple equations for the parameters and numerical estimates.  Finally, the summary of the results and the conclusions follow in Sec. \ref{sec:conclusions}.

\section{Dipolariton Hamiltonian}
\label{sec:dipolham}

\begin{figure}
\scalebox{0.65}{\includegraphics{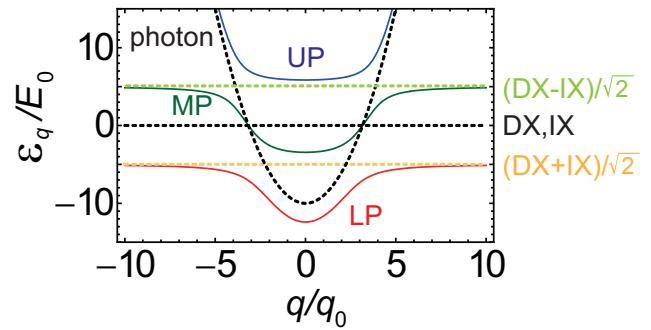}}
\caption{\label{fig:disp} (Color online) 
A typical dipolariton dispersion obtained by diagonalizing (\ref{singleparticleham}).  The LP, MP, and UP dispersions are the solid lines as marked.  For comparison we show the original DX, IX, and photon dispersions with no tunneling $ J = 0 $ and Rabi coupling $ \Omega = 0 $ (dashed lines as marked).  We also show 
the partially diagonalized dispersion including the tunneling but with $ \Omega = 0 $ (dashed lines as marked). The dispersion relation is measured in units of the experimental momentum scale $ q_0 $ and energy $ E_0 = \frac{\hbar^2 q_0^2}{2 m_{\mbox{\tiny ph}}} $. For a photon mass of $ 10^{-4} $ times the bare electron mass, and $ q_0 = 1 \mu m^{-1} $, $ E_0 = 0.38 $ meV.  The plot uses parameters $ \hbar \Omega/E_0 =10 $, $ \hbar J/E_0=10 $, $ \delta_{\mbox{\tiny IX}}/E_0=0 $, $ \delta_{\mbox{\tiny ph}}/E_0=-10 $, and $ M \gg m_{\mbox{\tiny ph}} $.  }
\end{figure}

A typical semiconductor microcavity system for dipolaritons is shown in Fig. \ref{fig:device}.  A coupled double quantum well is located between two sets of distributed Bragg reflectors (DBRs), forming a microcavity.  As is the case with standard exciton-polaritons, the microcavity allows for strong coupling between direct excitons 
in one of the quantum wells.  In addition to this, the barrier between the two quantum wells is made sufficiently thin such that tunneling may occur between them.  An applied bias voltage in the $z $-direction ensures that only the electron has the possibility of tunneling into the other quantum well. Due to the small effective mass of the electron, it may tunnel between the two quantum wells. On the other hand, the tunneling of the hole is negligible because of its larger effective mass and energy separation of hole levels in a coupled double quantum well \cite{cristofolini12}.  Thus there is a significant probability that the direct exciton may turn into an indirect exciton due to electron tunneling.  According to the parameters in Ref. \cite{cristofolini12}, the tunneling amplitude may be tunable to the same order of the Rabi splitting, thus a coherent superposition of the photon, direct exciton, and indirect exciton is a good approximation.  

Denoting the bosonic annihilation operators of the photon, DX, and IX as $  a  (\bm{R}),  e  (\bm{R}),  f  (\bm{R}) $ respectively, the total Hamiltonian of the system can be written
\begin{align}
\label{grandham}
H  & =  H_{\mbox{\tiny pol}} + H_{\mbox{\tiny int}} \\
H_{\mbox{\tiny pol}}  & = H_{\mbox{\tiny kin}} + H_{\mbox{\tiny Rabi}} + H_{\mbox{\tiny tun}} \label{polham}  .
\end{align}
The kinetic energy of the cavity photon, DX, and IX is
\begin{align}
H_{\mbox{\tiny kin}}   =  & \int d \bm{R} \Big(  a^\dagger (\bm{R}) {\cal H}_{\mbox{\tiny ph}}(\bm{R})  a (\bm{R})  +  e^\dagger (\bm{R}) {\cal H}_{\mbox{\tiny DX}}(\bm{R})  e (\bm{R}) \nonumber \\
& + f^\dagger (\bm{R}) {\cal H}_{\mbox{\tiny IX}}(\bm{R})  f (\bm{R}) \Big) 
\end{align}
where
\begin{align}
{\cal H}_{\mbox{\tiny ph}}(\bm{R}) & = - \frac{\hbar^2 }{2 m_{\mbox{\tiny ph}} } \nabla^2  +  \delta_{\mbox{\tiny ph}} ,\\
{\cal H}_{\mbox{\tiny DX}}(\bm{R}) & = - \frac{\hbar^2 }{2 M} \nabla^2 , \\
{\cal H}_{\mbox{\tiny IX}}(\bm{R}) & = - \frac{\hbar^2 }{2 M} \nabla^2   +  \delta_{\mbox{\tiny IX}} .
\end{align}
The zero energy point is taken to be the energy of the zero momentum $ q = 0 $ mode of the direct excitons.  The $ q = 0 $ modes of the photon and indirect exciton are taken to have 
a detuning of $ \delta_{\mbox{\tiny ph}} $ and $ \delta_{\mbox{\tiny IX}} $, respectively. The parameters involved in the above Hamiltonian are the exciton mass $ M = m_e+m_h $, where $ m_e $ ($m_h$) is the effective electron (hole) mass, and the photon effective mass is $ m_{\mbox{\tiny ph}} $. $ \bm{R} $ is the two dimensional center of mass position of the respective particles.  
 
The remaining single-particle terms are the Rabi coupling between the direct excitons and photons, 
\begin{align}
H_{\mbox{\tiny Rabi}}  & = -\frac{\hbar \Omega}{2} \sum_{\bm{q}} \left[ e_{\bm{q}}^\dagger a_{\bm{q}} + a_{\bm{q}}^\dagger e_{\bm{q}} \right] 
\end{align}
and the tunneling between the direct excitons and indirect excitons 
\begin{align}
H_{\mbox{\tiny tun}}  & = -\frac{\hbar J}{2} \sum_{\bm{q}}  \left[ e_{\bm{q}}^\dagger f_{\bm{q}} + f_{\bm{q}}^\dagger e_{\bm{q}} \right]  .
\end{align}
Here $ \hbar \Omega $ is the Rabi coupling photons and the DX, and the tunneling energy between DX and IX is $ \hbar J $.  The Fourier transforms are defined as 
$ e (\bm{R})  = \frac{1}{2 \pi } \sum_{\bm{q}} e^{i \bm{q} \cdot \bm{R} } e_{\bm{q}}, 
 f (\bm{R})  = \frac{1}{2 \pi } \sum_{\bm{q}} e^{i \bm{q} \cdot \bm{R} } f_{\bm{q}}, 
a (\bm{R})  = \frac{1}{2 \pi } \sum_{\bm{q}}  e^{i \bm{q} \cdot \bm{R} } a_{\bm{q}} $. 

Finally, the remaining term in (\ref{grandham}) is
\begin{align}
H_{\mbox{\tiny int}}  & =  H_{\mbox{\tiny DX-DX}} + H_{\mbox{\tiny IX-IX}} + H_{\mbox{\tiny IX-DX}} + H_{\mbox{\tiny sat}}
\end{align}
which are the non-linear interaction terms arising from DX-DX scattering ($H_{\mbox{\tiny DX-DX}}$), IX-IX scattering ($H_{\mbox{\tiny IX-IX}}$), IX-DX scattering ($H_{\mbox{\tiny IX-DX}}$), and a so-called ``saturation interaction'' ($ H_{\mbox{\tiny sat}} $) \cite{rochat00}  due to bosonization of the Rabi coupling.  
\begin{align}
H_{\mbox{\tiny DX-DX}} & = \frac{1}{2} \sum_{\bm{Q},\bm{Q}',\bm{q}} U_{\mbox{\tiny DX-DX}}(\bm{Q},\bm{Q}',\bm{q})   e^\dagger_{\bm{Q}-\bm{q}} e^\dagger_{\bm{Q}'+\bm{q}} e_{\bm{Q}'}  e_{\bm{Q}} , \\
H_{\mbox{\tiny IX-IX}} & = \frac{1}{2} \sum_{\bm{Q},\bm{Q}',\bm{q}} U_{\mbox{\tiny IX-IX}}(\bm{Q},\bm{Q}',\bm{q}) 
f^\dagger_{\bm{Q}-\bm{q}} f^\dagger_{\bm{Q}'+\bm{q}} f_{\bm{Q}'}  f_{\bm{Q}} , \\
H_{\mbox{\tiny IX-DX}} & = \sum_{\bm{Q},\bm{Q}',\bm{q}} U_{\mbox{\tiny IX-DX}}(\bm{Q},\bm{Q}',\bm{q}) 
e^\dagger_{\bm{Q}-\bm{q}} f^\dagger_{\bm{Q}'+\bm{q}} e_{\bm{Q}'}  f_{\bm{Q}} , \\
H_{\mbox{\tiny sat}}  & = \sum_{\bm{Q},\bm{Q}',\bm{q}} \left[ U_{\mbox{\tiny sat}} (\bm{Q},\bm{Q}',\bm{q})  a_{\bm{q}}^\dagger  b^\dagger_{\bm{Q}+\bm{Q}'-\bm{q}} 
b_{\bm{Q}'}  b_{\bm{Q}} + \mbox{H.c.} \right].  \label{satham}
\end{align}
We explain in more detail the origin of these terms and explicit expressions for the matrix elements $ U (\bm{Q},\bm{Q}',\bm{q}) $ in the following section.  

The non-interacting polariton Hamiltonian (\ref{polham}) may be diagonalized by the linear transformation
\begin{align}
\left(
\begin{array}{c}
p^{\mbox{\tiny LP}}_{\bm{q}} \\
p^{\mbox{\tiny MP}}_{\bm{q}} \\
p^{\mbox{\tiny UP}}_{\bm{q}} 
\end{array}
\right)
=
\left(
\begin{array}{ccc}
C^{\mbox{\tiny LP}}_{\bm{q}} & X^{\mbox{\tiny LP}}_{\bm{q}} & Y^{\mbox{\tiny LP}}_{\bm{q}}\\
C^{\mbox{\tiny MP}}_{\bm{q}} & X^{\mbox{\tiny MP}}_{\bm{q}} & Y^{\mbox{\tiny MP}}_{\bm{q}}\\
C^{\mbox{\tiny UP}}_{\bm{q}} & X^{\mbox{\tiny UP}}_{\bm{q}} & Y^{\mbox{\tiny UP}}_{\bm{q}}
\end{array}
\right)
\left(
\begin{array}{c}
a_{\bm{q}} \\
e_{\bm{q}}\\
f_{\bm{q}}
\end{array}
\right)
\label{transmatrix}
\end{align}
where $ C^{\mbox{\tiny LP,MP,UP}}_{\bm{k}},  X^{\mbox{\tiny LP,MP,UP}}_{\bm{k}},  Y^{\mbox{\tiny LP,MP,UP}}_{\bm{k}} $ are Hopfield coefficients for the 
photon, direct exciton, and indirect exciton components, respectively.  The new quasiparticles are lower polariton (LP), middle polariton (MP), and upper polaritons (UP).  The diagonalized non-interacting dipolariton Hamiltonian is
\begin{align}
H_{\mbox{\tiny pol}}  = \sum_{\bm{q}} \left[ \epsilon_{\bm{q}}^{\mbox{\tiny LP}} {p^{\mbox{\tiny LP}}_{\bm{q}}}^\dagger p^{\mbox{\tiny LP}}_{\bm{q}} 
+ \epsilon_{\bm{q}}^{\mbox{\tiny MP}} {p^{\mbox{\tiny MP}}_{\bm{q}}}^\dagger p^{\mbox{\tiny MP}}_{\bm{q}} 
+ \epsilon_{\bm{q}}^{\mbox{\tiny UP}} {p^{\mbox{\tiny UP}}_{\bm{q}}}^\dagger p^{\mbox{\tiny UP}}_{\bm{q}}  \right] ,
\end{align}
where $ \epsilon_{\bm{q}}^{\mbox{\tiny LP,MP,UP}} $ are the energy eigenvalues of the single particle Hamiltonian
\begin{align}
{\cal H}_{\mbox{\tiny pol}} = 
\left(
\begin{array}{ccc}
\frac{ \hbar^2 q^2}{2 m_{\mbox{\tiny ph}}}  + \delta_{\mbox{\tiny ph}} & -\frac{\hbar \Omega}{2}  & 0  \\
-\frac{\hbar  \Omega}{2}  & \frac{ \hbar^2 q^2}{2 M }   &  -\frac{\hbar J}{2}  \\
0 &  -\frac{\hbar J}{2} &  \frac{ \hbar^2 q^2}{2 M }  + \delta_{\mbox{\tiny IX}}
\end{array}
\right) . \label{singleparticleham}
\end{align}

A typical plot of the LP, MP, UP dispersions are shown in Fig. \ref{fig:disp}.  Although the three-way superposition makes 
the understanding of the dispersion more complicated than the simple anticrossing picture for exciton-polaritons, there is a simple way to understand the qualitative features of the spectrum.  First consider the DX and IX alone (i.e. $ \Omega = 0 $), and notice that due to the equality of the DX and IX mass the dispersions are separated by a constant amount for all $ q $. For $ \delta_{\mbox{\tiny IX}} = 0 $ the two quasiparticles are $ (e_{\bm{q}} \pm f_{\bm{q}})/\sqrt{2} $ with energies $ \mp J/2 $.  Now reinstating the photon coupling, we may think of the dipolariton as being a further admixture of the photon and the hybrid IX-DX particle.  For the case shown in Fig. \ref{fig:disp}, the LP dispersion is pushed down due to the anticrossing of the photon dispersion with respect to the $ (e_{\bm{q}} + f_{\bm{q}})/\sqrt{2} $ particle at energy $ -\hbar J/2 $.  This creates a typical LP dispersion similar to exciton-polaritons, but offset in energy by $ - \hbar J/2 $.  Thus, as far as the LP dispersion is concerned, the dipolariton dispersion shows the same essential features as standard exciton-polaritons.  

For sufficiently low temperatures $ k_B T < \hbar J $ we may expect that only the LP branch is populated, and we may ignore the MP and UP branches completely.  Dropping the ``LP'' labels in (\ref{transmatrix}), we may write an effective Hamiltonian only for the lower polaritons 
\begin{align}
H_{\mbox{\tiny LP}} = &  \sum_{\bm{q}} \epsilon_{\bm{q}} {p_{\bm{q}}}^\dagger p_{\bm{q}}  \nonumber \\
& +  \frac{1}{2} \sum_{\bm{Q},\bm{Q}',\bm{q}} U_{\mbox{\tiny LP}}(\bm{Q},\bm{Q}',\bm{q}) p^\dagger_{\bm{Q}-\bm{q}} p^\dagger_{\bm{Q}'+\bm{q}}  p_{\bm{Q}'}  p_{\bm{Q}} 
\label{dipolham}
\end{align}
where the effective LP interaction is
\begin{align}
U_{\mbox{\tiny LP}}(\bm{Q},\bm{Q}',\bm{q}) &  =  X_{\bm{Q}-\bm{q}}^* X_{\bm{Q}'+\bm{q}}^* X_{\bm{Q}} X_{\bm{Q}'}  U_{\mbox{\tiny DX-DX}}(\bm{Q},\bm{Q}',\bm{q})  \nonumber \\
& + Y_{\bm{Q}-\bm{q}}^* Y_{\bm{Q}'+\bm{q}}^* Y_{\bm{Q}} Y_{\bm{Q}'}  U_{\mbox{\tiny IX-IX}}(\bm{Q},\bm{Q}',\bm{q})  \nonumber \\
& + 2 X_{\bm{Q}-\bm{q}}^* Y_{\bm{Q}'+\bm{q}}^* Y_{\bm{Q}} X_{\bm{Q}'}  U_{\mbox{\tiny IX-DX}}(\bm{Q},\bm{Q}',\bm{q})  \nonumber \\
& + 2( C_{\bm{Q}-\bm{q}}^* X_{\bm{Q}'+\bm{q}}^* X_{\bm{Q}} X_{\bm{Q}'} + \mbox{H.c.})U_{\mbox{\tiny sat}}(\bm{Q},\bm{Q}',\bm{q})  .
\label{ulpexpression}
\end{align}
We see that the effective LP interaction consists of four parts, the mutual DX scattering, mutual IX scattering, IX-DX scattering, and the saturation interaction.  These terms are derived explicitly in the following section.

\section{Effective polariton interaction}
\label{sec:eff}

In this section we present the effective polariton interaction that consists from four terms: direct exciton-direct exciton (DX-DX), indirect exciton-indirect exciton (IX-IX), indirect exciton-direct exciton (IX-DX) interactions and the saturation interaction.

\subsection{DX-DX and IX-IX interaction}
\label{sec:interaction}

Calculation of the DX-DX and IX-IX interaction has already been performed in several previous works, hence we give a brief overview and restate the main results.  For the 
DX-DX interaction, an effective Hamiltonian in terms of excitons is derived from an electron-hole Hamiltonian involving Coulomb interactions \cite{ciuti98,deleon01,rochat00,hanamura79,stolz81,bobrysheva72}. The IX-IX interaction has been calculated
in Ref. \cite{byrnes10} by generalizing the methods of Refs. \cite{ciuti98,deleon01}.  While a variety of methods exist to derive the effective Hamiltonian, we follow the methods of de-Leon and Laikhtman \cite{deleon01} which gives a transparent and systematic way of obtaining the relevant quantities. We summarize the approach for the IX-IX interaction, which reduces to the DX-DX interaction by setting the interwell distance $ d $ to zero. 

The method starts with an antisymmetrized two IX wavefunction
\begin{align}
& \Phi_{\bm{Q} \bm{Q}'} (\bm{r}_e, \bm{r}_h, \bm{r}_{e'},\bm{r}_{h'} )= \frac{1}{\sqrt{2}} \Big( \frac{1}{\sqrt{2}}
\Big[ \Psi_{\bm{Q}} (\bm{r}_e, \bm{r}_h) \Psi_{\bm{Q}'} (\bm{r}_{e'}, \bm{r}_{h'})  \nonumber \\
& +  \Psi_{\bm{Q}} (\bm{r}_{e'}, \bm{r}_{h'})  \Psi_{\bm{Q}'}(\bm{r}_e, \bm{r}_h) \Big]
- \frac{1}{\sqrt{2}}  \Big[ \Psi_{\bm{Q}} (\bm{r}_{e'}, \bm{r}_h) \Psi_{\bm{Q}'} (\bm{r}_e, \bm{r}_{h'})  \nonumber \\
& + \Psi_{\bm{Q}} (\bm{r}_e, \bm{r}_{h'})  \Psi_{\bm{Q}'}(\bm{r}_{e'}, \bm{r}_h) \Big]  \Big)
\label{antisym}
\end{align}
where $ \Psi_{\bm{Q}} (\bm{r}_e, \bm{r}_h) $ is an IX wavefunction with center of mass momentum $ \bm{Q} $, taken as the two dimensional 
1s wavefunction \cite{leavitt90,byrnes10} in the quantum well plane and a delta-function in the $ z $-direction for the electrons and holes.  This assumes that for an IX the electron is always perfectly localized in the electron quantum well, and similarly for the holes.  Here $ \bm{r}_{e,h} $ are the three dimensional coordinates of the electrons and holes respectively. The Hamiltonian of the two exciton system is 
\begin{align}
{\cal H}  = & {\cal H}_0 + {\cal H}_1 \label{totham} \\
 {\cal H}_0  = &-\frac{\hbar^2}{2m_e} \nabla_e^2 -\frac{\hbar^2}{2m_h} \nabla_h^2 -\frac{\hbar^2}{2m_{e'}} \nabla_{e'}^2 
-\frac{\hbar^2}{2m_{h'}} \nabla_{h'}^2 \nonumber \\
& -V( | \bm{r}_e - \bm{r}_h|) -V( | \bm{r}_{e'} - \bm{r}_{h'}|) \nonumber \\
{\cal H}_1 = & V( | \bm{r}_e - \bm{r}_{e'}|)+ V( | \bm{r}_h - \bm{r}_{h'}|)  \nonumber \\
& -  V( | \bm{r}_e - \bm{r}_{h'}|) - V(| \bm{r}_h - \bm{r}_{e'}|)
\label{twoexcham}
\end{align}
with $ V(r) = e^2/ 4 \pi \epsilon r $ ($ \epsilon \approx 13 \epsilon_0 $ is the 
permittivity in GaAs, where $ \epsilon_0 $ is the permittivity in free space).  Taking the expectation value of (\ref{twoexcham}) with respect to $ \Phi_{\bm{Q} \bm{Q}'} $ and $ \Phi_{\bm{Q}+\bm{q} \bm{Q}'-\bm{q}} $ one obtains
the effective interaction \cite{byrnes10}
\begin{align}
& U_{\mbox{\tiny IX-IX}} (\bm{Q}, \bm{Q}',\bm{q})  = \frac{1}{A} \frac{e^2}{4 \pi \epsilon} 
a_B \left( \frac{2}{\pi} \right)^2 \Big[ I_{\mbox{\tiny dir}} (q,d) \nonumber \\
& + I_{\mbox{\tiny dir}} (\sqrt{(\Delta Q)^2 + q^2 - 2 \Delta Q q \cos \theta } ,d) \nonumber \\
& - I_{\mbox{\tiny exch}} ( \Delta Q,q,\theta,\beta_e,d) - I_{\mbox{\tiny exch}} ( \Delta Q,q,\theta,\beta_h,d) \Big]
\label{totalinteractionfull}
\end{align}
where $ a_B = 4 \pi \epsilon \hbar^2 / 2 e^2 \mu $
is the 2D Bohr radius, $ A $ is the trapping area of the excitons, and $ \mu $ is the reduced mass
$ \mu = m_e m_h /(m_e + m_h) $.  The dimensionless integrals $ I_{\mbox{\tiny dir}} $ and $ I_{\mbox{\tiny exch}} $ are given in Eqs. (A2) and (A5) and plotted in  Figs. 2 and 3 of Ref. \cite{byrnes10} respectively.  The DX-DX interaction is simply the same as this but evaluated at $ d =0 $ 
\begin{align}
U_{\mbox{\tiny DX-DX}} (\bm{Q}, \bm{Q}',\bm{q})  = U_{\mbox{\tiny IX-IX}} (\bm{Q}, \bm{Q}',\bm{q}) \left|_{d=0} \right.
\label{dxdxint}
\end{align}
Expressions for $ I_{\mbox{\tiny dir}} $ and $ I_{\mbox{\tiny exch}} $ are given in Eqs. (20) and (B1) and plotted in Figs. 1 and 2 of Ref. \cite{ciuti98} respectively. 

As discussed in Ref. \cite{byrnes10}, the most relevant momentum scale in (\ref{totalinteractionfull}) and (\ref{dxdxint}) is the $ Q, Q', q \rightarrow 0 $ limit since their characteristic momentum scale is of the order of $ \sim 1/a_B $ which is quite large compared to typical
experimental situations.  From Fig. 4 in Ref. \cite{byrnes10} to a good approximation the interaction is \footnote{We note that there is an error in Ref. \cite{byrnes10} due to the mismatch of the definition of Eq. (4) in this paper and the evaluated expressions Eq. (28) and (40).  In Eq. (4) the exciton exchange and the hole exchange terms are counted in the summation by the term $ \bm{q} = \bm{Q}'-\bm{Q} $.  Therefore these terms do not need to be included in Eq. (28) and (40), resulting in an extra factor of 2.}
\begin{align}
U_{\mbox{\tiny IX-IX}} (Q=0,Q'=0,q=0)  \approx   \frac{e^2}{4 \pi \epsilon a_B}  \frac{a_B^2}{A} \left( 6 + 3.5 \frac{d}{a_B} \right)  .
\label{ixixint}
\end{align}
Setting $ d=0 $ agrees with the standard estimate of $   \frac{6 e^2}{4 \pi \epsilon a_B}  \frac{a_B^2}{A} $ for the DX-DX interaction.

\subsection{Saturation interaction}
\label{sec:saturation}

Similar methods may be used to derive the ``saturation'' interaction, originating from corrections to the Rabi coupling due to the antisymmetrized two exciton wavefunction (\ref{antisym}).  Following Ref. \cite{byrnes10} we have 
\begin{equation}
U_{\mbox{\tiny sat}} (\bm{Q},\bm{Q}',\bm{q}) = \frac{\hbar \Omega}{2} \frac{a_B^2}{A} \sqrt{\frac{\pi}{2}} I_{\mbox{\tiny sat}} (\bm{Q},\bm{Q}',\bm{q}) .
\label{saturationintegral2}
\end{equation}
where $ I_{\mbox{\tiny sat}} (\bm{Q},\bm{Q}',\bm{q}) $ is given in Eq. (A8) and Fig. 3 in Ref. \cite{byrnes12}. This factor again has a characteristic momentum scale of $ \sim 1/a_B $, thus for typical experiments only the $ Q, Q', q \rightarrow 0 $ is of significance. We will therefore be interested in the value
\begin{equation}
U_{\mbox{\tiny sat}} (Q=0,Q'=0,q=0) \approx  3.5 \frac{\hbar \Omega}{2} \frac{a_B^2}{A}  .
\end{equation}

\begin{figure}
\scalebox{0.5}{\includegraphics{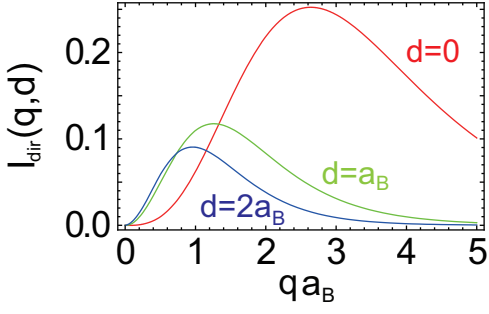}}
\caption{\label{fig:direct} (Color online) 
The direct IX-DX interaction integral $ I_{\mbox{\tiny dir}} (q,d) $ for three quantum well separations $ d $ as marked. The momentum $ q $ is the momentum transferred between the IX and DX. }
\end{figure}

\subsection{IX-DX Interaction}
\label{sec:ixdx}

The IX-DX interaction is calculated using analogous methods as described in Sec. \ref{sec:interaction}.  We start with a composite IX and DX wavefunction written as
\begin{align}
\phi_{\bm{Q} \bm{Q}'} (\bm{r}_e, \bm{r}_h, \bm{r}_{e'},\bm{r}_{h'} )= &  \frac{1}{\sqrt{2}} \Big[
\Psi_{\bm{Q}} (\bm{r}_e, \bm{r}_h) \psi_{\bm{Q}'} (\bm{r}_{e'}, \bm{r}_{h'})  \nonumber \\
& - \Psi_{\bm{Q}} (\bm{r}_{e'}, \bm{r}_h) \psi_{\bm{Q}'} (\bm{r}_{e}, \bm{r}_{h'})  \Big] .
\label{totalwave}
\end{align}
Here only the coordinates for the holes have been antisymmetrized as the electrons lie in different layers and are distinguishable.  In (\ref{totalwave}) $ \Psi_{\bm{Q}} $ denotes the IX wavefunction, and $ \psi_{\bm{Q}} $ is the DX wavefunction each with center of mass momentum $ \bm{Q} $. Taking the matrix element of (\ref{totham}) with $ \phi_{\bm{Q} \bm{Q}'} $ and $ \phi_{\bm{Q}+\bm{q} ~ \bm{Q}'-\bm{q}} $ we obtain the contributions 
\begin{align}
 U_{\mbox{\tiny IX-DX}}(\bm{Q},\bm{Q}',\bm{q})  = & U_{\mbox{\tiny dir}} (\bm{Q}, \bm{Q}',\bm{q}) + U_{\mbox{\tiny exch}} (\bm{Q}, \bm{Q}',\bm{q}) \nonumber \\
& + {\cal K} (\bm{Q}, \bm{Q}',\bm{q}) \label{totalixdx}
\end{align}
where
\begin{align}
& U_{\mbox{\tiny dir}} (\bm{Q}, \bm{Q}',\bm{q}) =  \int d \bm{r}_e d \bm{r}_h d \bm{r}_{e'} d \bm{r}_{h'}  \nonumber \\
& \Psi_{\bm{Q}}^* (\bm{r}_e, \bm{r}_h) \psi_{\bm{Q}'}^* (\bm{r}_{e'}, \bm{r}_{h'})  {\cal H} 
\Psi_{\bm{Q}+\bm{q}} (\bm{r}_e, \bm{r}_h) \psi_{\bm{Q}'-\bm{q}} (\bm{r}_{e'}, \bm{r}_{h'}) , \label{dirdefinition} \\
& U_{\mbox{\tiny exch}} (\bm{Q}, \bm{Q}',\bm{q}) =  -\int d \bm{r}_e d \bm{r}_h d \bm{r}_{e'} d \bm{r}_{h'}  \nonumber \\
& \Psi_{\bm{Q}}^* (\bm{r}_e, \bm{r}_{h'}) \psi_{\bm{Q}'}^* (\bm{r}_{e'}, \bm{r}_h)  {\cal H} 
\Psi_{\bm{Q}+\bm{q}} (\bm{r}_e, \bm{r}_h) \psi_{\bm{Q}'-\bm{q}} (\bm{r}_{e'}, \bm{r}_{h'}) .
\label{directexchdefinition}
\end{align}
The last term in (\ref{totalixdx}) is a correction term to take into account for the fact that the wavefunction (\ref{totalwave}) does not obey orthonormality, and gives spurious ``kinematic corrections'' \cite{deleon01,byrnes10}.  The correction factor is 
\begin{align}
& {\cal K} (\bm{Q}, \bm{Q}',\bm{q}) =  - \frac{1}{2} \int d \bm{r}_e d \bm{r}_h d \bm{r}_{e'} d \bm{r}_{h'} \Psi_{\bm{Q}}^* (\bm{r}_e, \bm{r}_h) \psi_{\bm{Q}'}^* (\bm{r}_{e'}, \bm{r}_{h'})   \nonumber \\
& \times \left( {\cal H}_0  {\cal A} + {\cal A}  {\cal H}_0  \right)
\Psi_{\bm{Q}} (\bm{r}_e, \bm{r}_h) \psi_{\bm{Q}'} (\bm{r}_{e'}, \bm{r}_{h'}) , 
\end{align}
with the non-orthonormality factor
\begin{align}
& {\cal A}(\bm{Q}, \bm{Q}',\bm{q}) = -  \int d \bm{r}_e d \bm{r}_h d \bm{r}_{e'} d \bm{r}_{h'} \nonumber \\
& \Psi_{\bm{Q}}^* (\bm{r}_{e}, \bm{r}_{h'}) \psi_{\bm{Q}'}^* (\bm{r}_{e'}, \bm{r}_{h} ) \Psi_{\bm{Q}+\bm{q}} (\bm{r}_{e}, \bm{r}_h) \psi_{\bm{Q}'-\bm{q}} (\bm{r}_{e'}, \bm{r}_{h'}) .
\end{align}

To evaluate the expressions above, we use an approximate form for the IX ground state wavefunction, as obtained in Ref. \cite{leavitt90}:
%
\begin{align}
\Psi_{\bm{Q}} (\bm{r}_e, \bm{r}_h) = &  \frac{1}{\sqrt{A}} 
e^{  i \bm{Q} \cdot \bm{R} } G(\rho,Z) \delta (z_e - d/2)  \delta (z_h + d/2)  ,
\label{approximateexciton}
\end{align}
where $ \rho = \sqrt{ (x_e-x_h)^2 + (y_e-y_h)^2 } $, $ Z = z_e-z_h $,  $ \beta_{e,h} = m_{e,h}/(m_e + m_h) $, and $ \bm{R} = \beta_e \bm{r}_e + \beta_h \bm{r}_h $. 
In Eq. (\ref{approximateexciton}), the exciton is considered to be trapped in a large area $ A $, such that the center of mass wavefunction is of the form of a plane wave.  We have assumed that in the $ z $ direction the electrons (holes) are completely confined to their respective quantum wells at position $ d/2 $ ($-d/2$), with delta function wavefunctions for simplicity. 
 The wavefunction that describes the state of the exciton for the relative coordinates is
%
\begin{align}
\label{excitonwavefunc}
G(\rho,Z) = & \frac{N_G}{a_B} \exp \Big[ - \frac{\lambda(Z)}{2}  \nonumber \\
& \times \Big( \sqrt{ (\rho/a_B)^2 + (Z/a_B)^2 } - Z/a_B \Big) \Big] ,
\end{align}
where $ \lambda(Z) = 2/(1+\sqrt{2 Z/a_B}) $ and $ N_G = \sqrt{\frac{\lambda(Z)^2}{2 \pi (1+  Z \lambda(Z)/a_B )}} $ is a normalization factor. The DX wavefunction can be obtained with the substitution $ Z = 0 $, giving
\begin{align}
\psi_{\bm{Q}} (\bm{r}_e, \bm{r}_h) = & \sqrt{\frac{2}{\pi a_B^2 A}} 
e^{ i \bm{Q} \cdot \bm{R} } e^{ - \rho/a_B }    \delta (z_e + d/2)  \delta (z_h + d/2)  .
\label{dxwave}
\end{align}

\begin{figure}
\scalebox{0.5}{\includegraphics{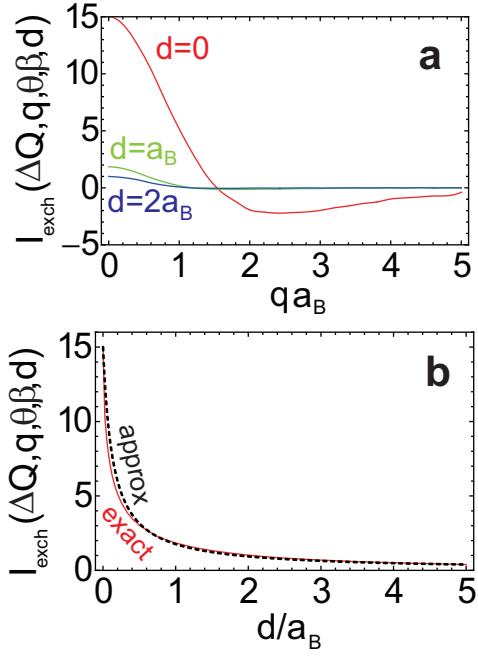}}
\caption{\label{fig:exchange} (Color online) 
The exchange IX-DX interaction integral $ I_{\mbox{\tiny exch}} (\Delta Q,q,\theta,\beta_e,d) $. 
(a) The transfer momentum $ q $ dependence for three quantum well separations $ d $ as marked for $ \Delta Q= 0 $. (b) The dependence on interwell distance $ d $ for $ \Delta Q= q= 0 $. Solid line shows the exact result and dashed line shows the approximation (\ref{approxiexch}).   Parameters used are for GaAs with $ \beta_e = 0.067/0.167=0.4 $. At $ \Delta Q=0 $ there is no dependence on $ \theta $.  
}
\end{figure}

The evaluations of the various terms are deferred to the Appendix \ref{app:effectiveinteraction}. Finally we find the direct term to be 
\begin{align}
\label{finaldirect}
U_{\mbox{\tiny dir}} (\bm{Q}, \bm{Q}',\bm{q}) = & \left[ -E_{1s}^{\mbox{\tiny IX}} -E_{1s}^{\mbox{\tiny DX}} + \frac{\hbar^2}{2M} (\bm{Q}^2 + \bm{Q}'^2) 
\right]\delta( \bm{q}) \nonumber \\
& +\frac{1}{A} \frac{e^2}{4 \pi \epsilon} a_B \left( \frac{2}{\pi} \right)^2 I_{\mbox{\tiny dir}} (q,d) .
\end{align}
where $ E_{1s}^{\mbox{\tiny IX,DX}} $ is the binding energy of a $1s$ exciton for the IX and DX, respectively. The function $ I_{\mbox{\tiny dir}} (q,d) $ is plotted for various $ d $ in Fig. \ref{fig:direct}. As is the case for the DX-DX interactions, in the limit of $ \bm{Q}, \bm{Q}',\bm{q} \rightarrow 0 $ this contribution becomes zero for all $ d $.  Thus the dominant contribution to the IX-DX interaction is due to the exchange term.  We evaluate the exchange term to be
\begin{align}
\label{holeexchange}
 U_{\mbox{\tiny exch}} &(\bm{Q}, \bm{Q}',\bm{q}) =  \nonumber \\
& \Big[ -E_{1s}^{\mbox{\tiny IX}} -E_{1s}^{\mbox{\tiny DX}} + \frac{\hbar^2}{2M} (\bm{Q}^2 + \bm{Q}'^2) \Big]  
{\cal A}(\bm{Q}, \bm{Q}',\bm{q})  \nonumber \\
& - \frac{1}{A} \frac{e^2}{4 \pi \epsilon} 
a_B \left( \frac{2}{\pi} \right)^2  I_{\mbox{\tiny exch}} ( \Delta Q,q,\theta,\beta_e,d) ,
\end{align}
where $ \Delta Q = |\bm{Q}'- \bm{Q}| $ and $ \theta $ is the angle between $ \bm{Q}'- \bm{Q} $ and $ \bm{q} $. The first term proportional to 
$ { \cal A }$ cancel with the corrections due to non-orthonormality (i.e. the last term in (\ref{totalixdx})). Numerical evaluations of the exchange integral $ I_{\mbox{\tiny exch}} $ are shown in Fig. \ref{fig:exchange}. We see that this term survives in the limit $ \bm{Q}, \bm{Q}',\bm{q} \rightarrow 0 $ and to a good approximation will give all of the contribution to the IX-DX scattering.  

Substituting (\ref{finaldirect}) and (\ref{holeexchange}) into (\ref{totalixdx}) we obtain the final effective Hamiltonian for the 
two-exciton system.  Subtracting the kinetic energy and binding energy terms, we obtain an expression for exciton-exciton interaction 
\begin{align}
U_{\mbox{\tiny IX-DX}} & (\bm{Q}, \bm{Q}',\bm{q})  = \nonumber \\
& \frac{1}{A} \frac{e^2}{4 \pi \epsilon} 
a_B \left( \frac{2}{\pi} \right)^2 \Big[ I_{\mbox{\tiny dir}} (q,d) - I_{\mbox{\tiny exch}} ( \Delta Q,q,\theta,\beta_e,d) \Big]
\end{align}
As before, the primary region of experimental interest is the  $ \bm{Q}, \bm{Q}',\bm{q} \rightarrow 0 $ limit.  To a good approximation the dimensionless integrals obey in this limit 
\begin{align}
I_{\mbox{\tiny dir}} ( q=0,d) & \approx  0, \nonumber \\
I_{\mbox{\tiny exch}} ( \Delta Q=0,q=0,\theta,\beta_e,d) & \approx \frac{1}{1/15 + d/2 a_B } . \label{approxiexch}
\end{align}
A comparison of the approximation (\ref{approxiexch}) with a numerical integration is shown in Fig. \ref{fig:exchange}.  We see that the approximation works rather well for the full range, and falls off fairly quickly for interwell distances of the order of the Bohr radius.  Our final expression is therefore
\begin{align}
U_{\mbox{\tiny IX-DX}}(Q=0,Q'=0,q=0)  \approx & 
\frac{e^2}{4 \pi \epsilon a_B}  \frac{a_B^2}{A}  \frac{1}{\frac{1}{6}+ 1.2\frac{ d}{a_B} } .
\label{approxixdx}
\end{align}

\section{Dissipative Gross-Pitaevskii equation for dipolaritons}
\label{sec:gpequation}

Eq. (\ref{dipolham}) gives an effective Hamiltonian that describes the low energy excitations of the dipolariton system.  The effective dipolariton-dipolariton interaction may be obtained by substituting the results of Sec. \ref{sec:eff} into (\ref{ulpexpression}).  
As with exciton-polaritons, what is of primary interest is the condensation of these particles.  We now describe the relevant parameters so that an effective theory of condensation of dipolaritons may be written. 

First let us evaluate the basic quantities of the dipolariton LP mass and lifetime.  This may be calculated by writing (\ref{singleparticleham}) as
\begin{align}
{\cal H}_{\mbox{\tiny pol}} & = {\cal H}_{q=0} + {\cal V}_{q} + \Gamma \nonumber \\
{\cal H}_{q=0}  & = \left(
\begin{array}{ccc}
\delta_{\mbox{\tiny ph}}  & -\frac{\hbar \Omega}{2}  & 0  \\
-\frac{\hbar  \Omega}{2}  & 0  &  -\frac{\hbar J}{2}  \\
0 &  -\frac{\hbar J}{2} &   \delta_{\mbox{\tiny IX}} 
\end{array}
\right)  ,\nonumber \\
 {\cal V}_{q}   & = \left(
\begin{array}{ccc}
\frac{ \hbar^2 q^2}{2 m_{\mbox{\tiny ph}}} & 0  & 0  \\
0 & \frac{ \hbar^2 q^2}{2 M }   &  0 \\
0 & 0 &  \frac{ \hbar^2 q^2}{2 M }  
\end{array}
\right) , \nonumber \\
\Gamma & = \frac{\hbar }{2} \left(
\begin{array}{ccc}
 - i \gamma_{\mbox{\tiny ph}}  &0  & 0  \\
0 & - i \gamma_{\mbox{\tiny DX}}   & 0  \\
0 &  0&   - i \gamma_{\mbox{\tiny IX}}
\end{array}
\right) ,
\end{align}
where the $ \Gamma $ contains the decay rates of each of the components related to the lifetimes by 
$ \tau_{\mbox{\tiny ph,DX,IX}} = 1/\gamma_{\mbox{\tiny ph,DX,IX}} $.  First treating $ \Gamma $ as a perturbation to $ {\cal H}_{q=0} + {\cal V}_{q}  $, we obtain the decay rate of the LPs
\begin{align}
\gamma_{\mbox{\tiny LP}} = \frac{1}{\tau_{\mbox{\tiny LP}}}  = \frac{|C_{\bm{q}}|^2}{\tau_{\mbox{\tiny ph}}} + \frac{|X_{\bm{q}}|^2}{\tau_{\mbox{\tiny DX}}}  + \frac{|Y_{\bm{q}}|^2}{\tau_{\mbox{\tiny IX}}} .
\end{align}
As with exciton-polaritons, for the case that  $ \tau_{\mbox{\tiny ph}} \ll \tau_{\mbox{\tiny DX}}, \tau_{\mbox{\tiny IX}} $ we have
$ \tau_{\mbox{\tiny LP}} \approx \tau_{\mbox{\tiny ph}}/|C_{\bm{q}}|^2 $.  Thus the dipolariton LP lifetime is of the order of the photon lifetime. The LP mass is obtained by treating $  {\cal V}_{q} $ as a perturbation to $  {\cal H}_{q=0} $, and ignoring $ \Gamma $ for simplicity.  The LP mass is
\begin{align}
\frac{1}{m_{\mbox{\tiny LP}}} = \frac{|C|^2}{m_{\mbox{\tiny ph}}} + \frac{|X|^2+ |Y|^2}{M},  
\label{effectivemasseq}
\end{align}
where for $ q=0 $ we have omitted the momentum labels on the Hopfield coefficients for brevity. 
For the typical case where $ m_{\mbox{\tiny ph}} \ll M $, we have $ m_{\mbox{\tiny LP}} \approx m_{\mbox{\tiny ph}}/|C|^2 $.  Again, the dipolariton LP mass is of the order of the photon effective mass. In the case of zero detuning $ \delta_{\mbox{\tiny ph}} = \delta_{\mbox{\tiny IX}}  = 0 $, the coefficients in (\ref{effectivemasseq}) are $|X|^2   = 1/2$ and 
$|C|^2 =  |Y|^2 = 1/4$.  Finally, for condensation of dipolaritons, the relevant interaction parameter is the low energy scattering $ Q, Q', q \rightarrow 0 $.  Compiling the results of Sec. \ref{sec:eff} and substituting this into (\ref{ulpexpression}), we obtain
\begin{align}
\hbar g = & \frac{e^2}{4 \pi \epsilon a_B} \frac{a_B^2}{A} 
\left[ 6 |X|^4 + (6+3.5 \frac{d}{a_B}) |Y|^4 + \frac{2|X|^2 |Y|^2}{\frac{1}{6}+ 1.2 \frac{d}{a_B} } \right]  \nonumber \\
& + 3.5 \hbar \Omega \frac{a_B^2}{A} |X|^2 (C^* X + X C^*) .
\label{mainresult}
\end{align}
Due to the relatively weak IX-DX interaction, most of the contribution will result from the IX-IX and 
the DX-DX interactions.  Thus the dipolariton-dipolariton interactions are generally of the same order as those for standard exciton-polaritons. 

We are now in a position to write down an equation which describes the dipolariton condensate.  Due to the close similarity of the physics of dipolaritons to exciton-polaritons, we may assume that condensation of dipolaritons also occurs to form a macroscopically occupied ground state \cite{su14}.  For details of exciton-polariton condensation see review articles such as Refs. \cite{deng10,keeling11,carusotto13}.  Incoherent pumping of the dipolariton system results in initially a large population of reservoir excitons, corresponding to DX or IX depending on the pumping scheme. These excitons cool within the semiconductor via phonon emission, up to a bottleneck momentum, where the photon fraction becomes appreciable.  A bottleneck population at momentum such that $ \frac{\hbar^2 k^2}{2 m_{\mbox{\tiny ph}}} \sim \hbar \Omega $ is then created, after which dipolariton-dipolariton scattering becomes the dominant mechanism of dipolariton momentum transfer.  From the results of Sec. \ref{sec:eff}, the magnitude of the dipolariton-dipolariton scattering is of the same order as for standard excitons, hence this should occur rather efficiently.  

At sufficiently high reservoir densities, a macroscopic population of dipolaritons should form at $ k = 0 $.
For non-zero temperatures, $T > 0$, there is no condensate in an infinite two-dimensional  system. However,
condensation is possible in a finite-sized system \cite{bagnatto91,nozieres95,dalfovo99}. 
A trapped, ideal Bose gas undergoes a transition to a condensed state at the critical  temperature 
$k_B T_c = 12 \hbar^2  n / \pi m s$, 
where $s=2$ is the spin degeneracy,  $n$ is the polariton density in a trap \cite{bagnatto91,berman06}. For lower polaritons with density $n \sim 10^9$ cm$^{-2}$, one obtains $T_c \sim 170$ K.

For exciton-polaritons, the dissipative Gross-Pitaevskii (GP) equation captures the condensate dynamics to a good approximation \cite{wouters07,carusotto13}.  For low enough temperatures below $ T_c $ we may thus also write for dipolaritons
\begin{align}
i \frac{\partial \varphi( \bm{R})}{\partial t} = & \Big[ \frac{\hbar \nabla^2}{2 m_{\mbox{\tiny LP}}} + \frac{ V( \bm{R})}{\hbar} + \frac{i}{2}\left[ {\cal R}(n ( \bm{R}) ) - \gamma_{\mbox{\tiny LP}} \right] \nonumber \\
& + g | \varphi ( \bm{R})  |^2 + 2 g n ( \bm{R}) \Big] \varphi  ( \bm{R})
\end{align}
and the reservoir obeys
\begin{align}
\frac{\partial n( \bm{R})}{\partial t}  = P - \gamma_R n( \bm{R}) -  {\cal R}(n ( \bm{R})) | \varphi ( \bm{R}) |^2 . 
\end{align}
Here $ \varphi (\bm{R}) $ is the macroscopic wavefunction of a dipolariton condensate, $ V( \bm{R}) $ is the spatial trapping potential, $ n (\bm{R})  $ is the reservoir density, $ P $ is the pumping rate of the reservoir, $ \gamma_R $ is the decay rate of the reservoir, and $ R $ is the stimulated scattering of the reservoir excitons into the $ k = 0 $ dipolariton mode.  The only remaining unspecified parameters in the dissipative GP-equation are the reservoir-condensate scattering $ {\cal R} $ and the pumping rate $ P $.  Even for the exciton-polariton case these two parameters are typically put in phenomenologically, due to the difficulty of precisely modeling the reservoir.  However, we do know that the scattering occurs due to interactions of an incoherent reservoir of DX and IX, which was calculated in Sec. \ref{sec:eff} to be very similar to the DX scattering for the exciton-polariton case \cite{wouters07}.  Therefore it is reasonable to assume similar values to that used for exciton-polaritons, as with the phenomenological pump rate $ P $.

\section{Summary and Conclusions}
\label{sec:conclusions}

We have obtained a simple theoretical description of dipolaritons in a coupled double well microcavity system. Overall, the effective parameters as derived in Sec. \ref{sec:gpequation} suggest that the admixture of the indirect excitons give only a minor modification of the essential parameters.  As with exciton-polaritons, the effective dipolariton mass is of the order of the light photon effective mass, and the lifetime is 
of the order of the photon lifetime.  The dipolariton LP dispersion shows the same general behavior with a sharp anticrossing at momenta  $ \frac{\hbar^2 k^2}{2 m_{\mbox{\tiny ph}}} \sim \hbar \Omega $. The effective dipolariton-dipolariton interaction was calculated 
and was found to be expressible by the simple relation Eq. (\ref{mainresult}).  There are four contributions to the effective interaction, resulting from the DX-DX, IX-IX, IX-DX scattering, and the saturation interaction.  The relative strength of these contributions depend upon the photon, DX, and IX fractions, which may be tuned by changing the detuning and tunneling strength $ J $.  Under typical parameters where each of the fractions are comparable, the dominant effective interaction originates from the IX-IX interaction and the DX-DX interaction.  Due to the fast fall-off of the IX-DX contribution with the interwell distance $ d $ (Fig. \ref{fig:exchange}(b)), this contribution is typically the weakest of the four. 

Due to the dipolar nature of the dipolaritons, one may expect that the IX-IX would be considerably stronger than the DX-DX interactions, which originate from purely an exchange effect.  However, as may be observed from (\ref{ixixint}) the increase in interaction is relatively weak, only linearly increasing with the interwell separation $ d $.  This would suggest that contrary to expectation, the range of tunability of the dipolaritons is only moderate compared to exciton-polaritons, which may also be tuned by varying the photon and exciton fractions. However, the similar parameters and dispersion characteristics suggest that there should be no impediment in principle for condensation of dipolaritons, with similar physics to exciton-polaritons.  Similar configurations may be possible with a polaritons formed by a coupled double graphene layer in a microcavity structure \cite{berman10,berman12}. Even without the feature of tunability, this would open a fascinating variant of the exciton-polariton condensate both from a fundamental point of view and technological applications.

\section*{Acknowledgements}

This work is supported by the Transdisciplinary Research Integration Center, the Okawa foundation, the Inamori foundation, NTT, and JSPS KAKENHI Grant Number 26790061.  G.K. is grateful to  Professional Staff Congress -- City University of New York for support, award \#66140-00~44.

\appendix

\section{Evaluation of DX-IX scattering integrals}

\label{app:effectiveinteraction}

Eq. (\ref{dirdefinition}) may be evaluated by making a change of variables to $ \bm{R}= \beta_e \bm{r}_e + \beta_h \bm{r}_h $ and 
$ \bm{\rho}= \bm{r}_e -  \bm{r}_h $, after which we obtain
%
\begin{align}
\label{directtermone}
& U_{\mbox{\tiny dir}} (\bm{Q}, \bm{Q}',\bm{q}) = 
\left[ -E_{1s}^{\mbox{\tiny IX}}  -E_{1s}^{\mbox{\tiny IX}} + \frac{\hbar^2}{2M} (\bm{Q}^2 + \bm{Q}'^2) 
\right]\delta( \bm{q})   \nonumber \\
& + \frac{e^2}{4 \pi \epsilon A} 
\int d^2 \rho d^2 \rho'  \frac{2 \pi}{q} | G(\rho, d) |^2 | G(\rho', 0) |^2 \nonumber \\
& \times  \big[  e^{-i \bm{q} \cdot \beta_h ( \bm{\rho} - \bm{\rho}' ) }e^{-dq} + 
e^{i \bm{q} \cdot \beta_e ( \bm{\rho} - \bm{\rho}' ) } \nonumber \\
& - e^{- i \bm{q} \cdot ( \beta_h \bm{\rho} +  \beta_e \bm{\rho}' ) }e^{-dq} - e^{i \bm{q} \cdot ( \beta_e \bm{\rho} +  \beta_h \bm{\rho}' ) }
\big] 
\end{align} 
Eq. (\ref{finaldirect}) may be obtained by performing the $ \rho $ and $ \rho' $ integrals separately and using the rotational invariance of $ \bm{q} $. 
Fig. \ref{fig:direct} is obtained by evaluating
\begin{align}
& I_{\mbox{\tiny dir}} (q,d) = \frac{2 \pi^5}{q a_B} \Big[  I_0^{\mbox{\tiny IX}} (q\beta_h) I_0^{\mbox{\tiny DX}} (q\beta_h) e^{-dq} 
 + I_0^{\mbox{\tiny IX}} (q\beta_e) I_0^{\mbox{\tiny DX}} (q\beta_e) \nonumber \\
& - I_0^{\mbox{\tiny IX}} (q\beta_h) I_0^{\mbox{\tiny DX}} (q\beta_e) e^{-dq} 
- I_0^{\mbox{\tiny IX}} (q\beta_e) I_0^{\mbox{\tiny DX}} (q\beta_h) \Big],
\end{align}
where
\begin{align}
I_0^{\mbox{\tiny IX}} (q) =  \int d \rho \rho J_0 ( q \rho) | G(\rho,d) |^2 
\end{align}
and
\begin{align}
I_0^{\mbox{\tiny DX}} (q) = &  \int d\rho \rho J_0 ( q \rho) | G(\rho,0) |^2  \nonumber\\
= & \frac{1}{2\pi} \frac{1}{(1 + (q a_B/2)^2 )^{3/2}}
\end{align}
and $ J_0 ( x) $ is the Bessel function of the first kind.  

The exchange integral may be obtained by following the derivation given in the Appendix B of Ref. 
\cite{ciuti98}. We obtain (\ref{holeexchange}) with
\begin{widetext}
\begin{align}
I_{\mbox{\tiny exch}} & ( \Delta Q,q,\theta,\beta,d) = 
\left( \frac{\pi}{2} \right)^2 \int_0^\infty dx \int_0^{2\pi} d \theta_x
\int_0^\infty dy_1 \int_0^{2\pi} d \theta_1 \int_0^\infty dy_2 \int_0^{2\pi} d \theta_2 
x y_1 y_2 \nonumber \\
& \times \cos \{ \Delta Q a_B [ \beta x \cos (\theta-\theta_x) + \beta y_1 \cos (\theta-\theta_1) ]  + q a_B [ -x \cos \theta_x -\beta y_1  \cos \theta_1 + (1-\beta) y_2
\cos \theta_2] \}  \nonumber \\
& \times G( a_B \sqrt{(y_2 \cos \theta_2 - y_1 \cos \theta_1 - x \cos \theta_x)^2 + (y_2 \sin \theta_2 - y_1 \sin \theta_1 - x \sin \theta_x)^2 }, 0)  G( a_B x, d)  G( a_B y_1, 0) G( a_B y_2, d) \nonumber \\
&  \Big[ \frac{1}{\sqrt{y_1^2 + x^2 + 2 y_1 x \cos ( \theta_1 - \theta_x)+ (d/a_B)^2}} 
+ \frac{1}{\sqrt{y_2^2 + x^2 - 2 y_2 x \cos ( \theta_2 - \theta_x)}} -\frac{1}{\sqrt{y_2^2 + (d/a_B)^2}} -\frac{1}{y_1}
\Big] .
\label{iexchexpression}
\end{align}
\end{widetext}


\end{document}